\documentclass[12pt]{article}
\usepackage{amssymb,amsmath,amsfonts,amssymb}
\usepackage{cite}

\newlength{\dinwidth}
\newlength{\dinmargin}
\def\cM{\mathcal{M}}
\def\cS{\mathcal{S}}
\def\cW{\mathcal{W}}

\def\RR{{\mathbb R}}

\def\ch{\mbox{ch}}
\def\sh{\mbox{sh}}

\newcommand{\wb}[1]{{\langle \, #1 \, \rangle_{\beta}}}
\newcommand{\wB}[1]{{\langle \, #1 \, \rangle_B}}
\newcommand{\w}[1]{{\langle \, #1 \, \rangle }}

\def\phiM{\phi_{{(\mathcal M},g)}}

\def\ie{\textit{i.e.\ }}

\begin{document}
\title{Local Temperature  in Curved Spacetime}
\author{Detlev~Buchholz \ and \ Jan~Schlemmer \\
Institut f\"ur Theoretische Physik,
Universit\"at G\"ottingen,  \\ 37077 G\"ottingen, Germany}
\date{}
\maketitle
\begin{abstract} \noindent
A physically meaningful local concept of temperature is
introduced in quantum field theory on curved spacetime and applied 
to the example of a massless field on de Sitter space.  
It turns out in this model that the equilibrium (Gibbs) 
states which can be prepared by a geodesic 
observer have in general a varying temperature distribution in 
the neighborhood of the geodesic and may not even allow for 
a consistent thermal interpretation close to the horizon. 
This result, which can be traced back to the Unruh effect, 
illustrates the failure of a global notion 
of temperature in curved spacetime and reveals the need 
for a local concept, as presented here. \\[2mm]
\end{abstract}
%\keywords{}
%\maketitle
%
\noindent The unambiguous 
determination of the temperature of physical systems 
in curved spacetime is an intricate problem since 
the measurement of temperature is obscured by 
effects of the underlying geometry. The most prominent phenomenon 
in this respect is the Unruh effect \cite{Un}, \ie the thermal 
response of measuring devices to accelerating 
(tidal) forces. In view of the 
omnipresence of such effects, it does not seem meaningful to rely 
on a global notion of temperature in curved spacetime. Instead, 
one needs a local definition  which can be applied to states 
of physical interest and which provides the desired 
information about their thermal properties.

In the present letter we propose such an operationally 
meaningful local concept of temperature 
in the general framework of quantum field theory  
on curved spacetimes \cite{Wa}. It is based on ideas expounded in 
\cite{BuOjRo,Bu} in the case of Minkowski space theories. 
These ideas can be 
carried over to quantum field theory on arbitrary spacetime
mani\-folds by making use of the notion of generally covariant quantum 
fields, introduced in \cite{BrFrVe}. 

Heuristically speaking, the latter approach provides a comprehensive
description of a given (Lagrangian) quantum field theory,    
simultaneously on all globally 
hyperbolic spacetime manifolds ${\cM}$ with Lorentzian metric $g$. 
A generally covariant (fundamental or composite) quantum field $\phi$
may be thought off as a master field 
whose restriction $\phiM$ to any spacetime $(\cM,g)$ defines a 
concrete field operator 
\begin{equation*}
\phiM(x) \, , \quad x \in \cM \, .
\end{equation*}
A tight relationship
between the field operators on different spacetimes is established 
by the condition of general covariance, \ie the requirement  
that the restrictions of the master field $\phi$ to 
isomorphic spacetimes are related by algebraic 
isomorphisms \cite{BrFrVe}. This novel framework has been crucial
for the development of renormalized perturbation theory on
curved spacetime \cite{HoWa} and, in particular, for a  
generally covariant definition of composite field operators,
such as the stress tensor \cite{HoWa2,Mo}.

We propose (cf.\ also \cite{Bu,Oj}) 
to use these master fields for the determination of the
local thermal properties of physical 
states on a given spacetime manifold 
$(\cM,g)$ by comparing them locally 
with global equilibrium states on Minkowski space $(\cM_0,g_0)$.

On Minkowski space, the global equilibrium states 
in a given Lorentz system are obtained as thermodynamic limits 
of Gibbs ensembles; alternatively, they can be
identified by the KMS--condition \cite{Ha}. The equilibrium states 
are parametrized by the inverse temperature $\beta = T^{-1}$ and 
possibly by other 
data such as chemical potentials or phase types. 
In order to simplify the discussion, we assume that we 
are dealing here with a theory having, for each $\beta > 0$,
a unique equilibrium state (expectation functional)
$\wb{\cdot}$ on Minkowski space $(\cM_0,g_0)$.
These equilibrium states are then necessarily 
stationary, homogeneous and isotropic.
In particular, the expectation values 
\begin{equation*}
\wb{\phi_{(\cM_0,g_0)} (x)} \, , \quad x \in {\cM_0} \, ,
\end{equation*}
do not depend on the spacetime point $x$; we 
therefore put $x = 0$ in the following.
In the present context it is of importance that each function 
\begin{equation*}
\beta \mapsto \Phi(\beta) \doteq
\wb{\phi_{(\cM_0,g_0)} (0)}
\end{equation*}
provides information about the temperature dependence of the
spacetime mean
\begin{equation*}
\Phi \doteq \lim_{V \nearrow 
\RR^4} \text{\small $\frac{1}{|V|}$} \int_V \! d^4 x \, \phi_{(\cM_0,g_0)}(x), 
\end{equation*}
which is some intensive macroscopic observable.  In fact, for any
mixture $\wB{\cdot} \doteq \int \! d\rho(\beta) \, \wb{\cdot}$
of equilibrium states, where $\rho$ is some probability measure 
describing possible temperature fluctuations, one has
\begin{equation} \label{f1} 
\wB{\phi_{(\cM_0,g_0)} (0)} 
= \! \int \! d\rho(\beta) \, \wb {\phi_{(\cM_0,g_0)} (0)} =
\int \! d\rho(\beta) \, \Phi(\beta) = \wB{\Phi} \, .
\end{equation}

States which are only 
locally in equilibrium are,
according to the ideas 
expounded in \cite{BuOjRo,Bu},  
expected to look like 
(mixtures of) global equilibrium states 
with regard to local measurements within   
a given set ${\cS}$ of suitable observable master fields $\phi$.
The proper choice of these ``local thermal observables'' 
requires some care, 
as was discussed in \cite{BuOjRo}: one may incorporate into $\cS$ 
normal products of master fields and their so--called balanced derivatives.
It should be noticed that 
familiar observables, such as the stress energy tensor, 
are not of this type and may therefore 
not be included into ${\cS}$. Nevertheless,
this restricted set of observables provides 
sufficient information in order to determine the thermal
properties of local equilibrium states \cite{BuOjRo,Bu}.   
An independent justification of the choice of ${\cS}$
proposed in \cite{BuOjRo} 
was recently provided in \cite{Sch2}. There it was shown 
that these observables can be understood as 
idealized measurements modelled 
by Unruh--de Witt detectors.  

After these preparations, we can characterize now 
those states on a given spacetime $(\cM,g)$
which are locally in equilibrium.

\noindent \textbf{Criterion:} Let $\w{\cdot}$ be a state on $(\cM,g)$.
This state is in equilibrium at  
spacetime point $x \in {\cM}$
with regard to the observables in ${\cS}$ if there 
exists some mixture of equilibrium states  $\w{\cdot}_B$
on Minkowski space  $(\cM_0,g_0)$ such that 
\begin{equation} \label{f2}
\w{\phi_{(\cM,g)} (x)} = \w{\phi_{(\cM_0,g_0)} (0)}_B \, , 
\quad \phi \in {\cS}.
\end{equation}
In more physical terms, there is no way to distinguish 
the given state  at 
the spacetime point $x$ from a global equilibrium situation.
Note that the reference state $\w{\cdot}_B$ may depend on $x$, 
thereby covering situations where the 
thermal properties of $\w{\cdot}$ vary from point to point. 

\vspace*{2mm}
Combining relations  (\ref{f1}) and (\ref{f2}), one arrives at 
the fundamental equality 
\begin{equation} \label{f3}
\w{\phi_{(\cM,g)} (x)} = 
\int \! d\rho_x(\beta) \, \Phi(\beta) \, , \quad \phi \in  {\cS} \, .
\end{equation}
It shows that one may reinterpret the expectation values 
of the microscopic observables $\phi \in {\cS}$ in local equilibrium states
on $(\cM,g)$ as a (in general $x$--dependent) mean of the
corresponding macroscopic observables $\Phi$. This equality
thus provides a link between the microscopic theory and its
macroscopic thermal interpretation. It has been applied 
in  \cite{BuOjRo,Bu,Ba} to local equilibrium states on Minkowski 
space, thereby providing interesting insights about the 
possible spacetime patterns of such states. 

In the present
letter, we apply this formalism to the simple example of a free massless 
scalar field on curved spacetime. The corresponding master field $\phi$
satisfies the field equation
\begin{equation} \label{f4}
\big( \square + \text{\small $\frac{R}{6}$} ) \, \phi_{(\cM,g)} (x) = 0 \, ,
\end{equation}
where $\square$ denotes the d`Alembertian 
and $R$ the scalar curvature of the respective 
spacetime ${(\cM,g)}$. The fields at different
spacetime points $x,y \in {\cM}$ satisfy the commutation relations
\begin{equation*} 
[\phi_{(\cM,g)}(x), \phi_{(\cM,g)}(y)] =
i \, \Delta_{(\cM,g)}(x,y) \, ,
\end{equation*}
where $\Delta_{(\cM,g)}$ denotes the commutator function, \ie the 
difference of the advanced and retarded fundamental solutions of 
equation (\ref{f4}) \cite{Wa}. Note that these relations are  
invariant under the global gauge transformation $\phi \mapsto - \phi$.

For the test of the thermal properties of the states 
in this model, it seems natural to rely on densities
which can be built out of products of the master field  $\phi$. 
The simplest example is the ``Wick square'' $\phi^2$
which is fixed, up to some curvature term,  
by a point splitting procedure consistent with the condition
of general covariance \cite{BrFrVe,HoWa3},
\begin{equation} \label{f5}
\hspace*{3mm} \phi^{\, 2}_{(\cM,g)}(x) = 
 \lim_{y \rightarrow x} \big(
\phi_{(\cM,g)}(x) \, \phi_{(\cM,g)}(y) -
H_{(\cM,g)}(x,y) \big) + cR  \, . 
\end{equation}
Here $H_{(\cM,g)}$ is the Hadamard--parametrix \cite{KaWa,Wa} resulting from 
equation (\ref{f4}) and $c$ is some \textit{a priori}
arbitrary constant which will be fixed below. 

Let us consider first the case of
four--dimensional Minkowski space $(\cM_0,g_0)$.
There $R = 0$ and the commutator function $\Delta_{(\cM_0,g_0)}$
is given in proper coordinates by
\begin{equation*} 
\Delta_{(\cM_0,g_0)}(x,y) = 
\text{\small $\frac{-1}{2 \pi}$} \, \varepsilon(x_0 - y_0 ) \,
\delta((x-y)^2) \, ,
\end{equation*}
where $\varepsilon$ denotes the sign function of the time
difference  $x_0 - y_0$ and $\delta$ the Dirac delta--function of the
Lorentz square $(x-y)^2$. 
The Hadamard--parametrix coincides in this case with the 
two--point function of the field in the 
Minkowskian vacuum state and is given by
\begin{equation} \label{f6}
H_{(\cM_0,g_0)}(x,y) = 
\text{\small $\frac{-1}{4\pi^2 \, (x - y -i0)^2}$} \, .
%\frac{-1}{4\pi \, (x - y -i0)^2} \, .
\end{equation}
Thus $ \phi^{\, 2}_{(\cM_0,g_0)}$ coincides 
with the familiar Wick square of the massless scalar free field.

The global, gauge--invariant and clustering (\ie primary)
equilibrium states of geo\-desic
observers on $(\cM_0,g_0)$ can easily be 
determined from the commutator function with the help of the KMS--condition
\cite{Ha}. For given $\beta$, their respective two--point functions are   
\begin{equation} \label{f7}
\wb{\phi_{(\cM_0,g_0)} (x)  \,  \phi_{(\cM_0,g_0)} (y)} = 
\text{\small $\frac{i}{2\pi}$}  \int \! d\omega \,
 \text{\small $\frac{1}{1 - e^{-\beta \omega}}$} \,
\int \! dt \, \Delta_{(\cM_0,g_0)}(x(t),y) \, e^{it\omega} \, ,
\end{equation}
where $t \mapsto x(t) = x + t e$ is the time evolution 
of $x$ in the time direction $e$ of the observer.
Noticing that from relation (\ref{f7}) one recovers the 
Hadamard--parametrix (\ref{f6}) in the zero--temperature
limit $\beta \rightarrow \infty$, one obtains
\begin{equation} \label{f8}
\begin{split}
& \wb{\big(\phi_{(\cM_0,g_0)} (x)  \,  \phi_{(\cM_0,g_0)} (y)
- H_{(\cM_0,g_0)}(x,y) \big) } = \\
& \hspace*{-2mm} 
\text{\small $\frac{i}{2 \pi}$} \! \int \! \! d\omega \,
 \big( \text{\small $\frac{1}{1 - e^{-\beta \omega}}$}
- \theta(\omega)\big) \! \!
\int \! dt \, \Delta_{(\cM_0,g_0)}(x(t),y) \, e^{it\omega} \, .
\end{split}
\end{equation}
In the latter relation, one can proceed to the limit
$y \rightarrow x$, giving after a straightforward computation
\begin{equation}  \label{f9}
\wb{\phi^{\, 2}_{(\cM_0,g_0)}(x)} =
 \text{\small $\frac{1}{2\pi^2}$} \! \int_0^\infty \! \! d\omega \,
 \text{\small $\frac{\omega}{e^{\beta \omega} - 1}$} 
=  \text{\small $\frac{1}{12\, \beta^2}$} \, .
\end{equation}
Thus we find that the intensive macroscopic observable corresponding
to $\phi^2$ is proportional to the square of the temperature.  
In a similar manner, one can con\-struct other densities and determine
their macroscopic thermal interpretation \cite{BuOjRo,Bu}. 

We turn now to a simple but physically significant example of
curved spacetime: four--dimensional de Sitter space. It can 
conveniently be represented in the ambient five--dimensional Minkowski space 
as the one--sheeted hyperboloid
\begin{equation*}
{\cM} \doteq \{ x  \in \RR^5 : x^2 = - r^2 \} 
\end{equation*}
with the induced metric $g$. Here $r$ is its radius, $R = 12 \, r^{-2}$ 
its scalar curvature and $\Lambda = 3 \, r^{-2}$ the corresponding
cosmological constant. The commutator function on $({\cM},g)$ is 
given by \cite{BrMo}
\begin{equation*}
\Delta_{({\cM},g)}(x,y) = 
\text{\small $\frac{-1}{2 \pi}$} \, \varepsilon(x_0 - y_0 ) \,
\delta((x-y)^2) \, ,
\end{equation*}
and the relevant part of the Hadamard--parametrix by 
\begin{equation*} 
H_{(\cM,g)}(x,y) = 
\text{\small $\frac{-1}{4\pi^2 \, (x - y -i0)^2} + \frac{1}{768 
    \pi^2}$} \, R \, , 
\end{equation*}
disregarding a term which vanishes  for $y \rightarrow x$.  
These expressions are to be understood as 
restrictions of the given distributions in five--dimensional
ambient Minkowski space to the four--dimensional manifold ${\cM}$.  
The first term in the above expression for the  
(truncated) Hadamard--parametrix has also here direct physical significance:
It is the two--point function of the Gibbons--Hawking state
\cite{GiHa} which is distinguished by the fact that it
is invariant under the action of the de Sitter group and
in thermal equilibrium for all geodesic observers at 
inverse temperature $\beta_r = T_r^{-1} = 2\pi r $ 
\cite{NaPeTh,BoBu}. 

In contrast to Minkowski space, one can assign to geodesic 
observers on de Sitter space only locally a dynamics 
(corresponding to a timelike Killing vectorfield). Without restriction
of generality, we consider in the following
an observer whose geodesic passes through the 
point $w = (0,r,0,0,0) \in \cM$ and is given, as a function of proper time, by 
$t \mapsto w(t) = (\sh(\text{\small $\frac{t}{r}$}) \, r, 
\ch(\text{\small $\frac{t}{r}$}) \, r, 0, 0, 0)$.
Neighboring points $x \in {\cM}$, 
which may be thought of as being
occupied by measuring devices dragged along by the 
observer, evolve according to
\begin{equation} \label{f10}
\hspace*{2mm} t \mapsto x(t) = 
\big(\ch(\text{\small $\frac{t}{r}$}) \, \! x_0 \! + \! 
\sh(\text{\small $\frac{t}{r}$}) \, \! x_1  , 
\ch(\text{\small $\frac{t}{r}$}) \, \! x_1 \! +  \!
\sh(\text{\small $\frac{t}{r}$}) \, \! x_0 ,x_2,x_3,x_4 \big) \, .
\end{equation}
These orbits are positive timelike only in the wedge--shaped 
region ${\cW} = \{ y \in {\cM} : y_1 > |y_0| \} $, 
which is the causal completion of the geodesic of
the observer and bounded by a bifurcate event horizon. 
The observer can prepare in the interior of his wedge ${\cW}$, 
for all positive $\beta$, gauge invariant and clustering 
KMS--states. They are, with regard to his 
dynamics, stationary, stable and passive in the sense that he 
cannot extract energy from them by operating a cyclic engine \cite{Ha}.
It would, however, be premature to visualize these states 
in analogy to global equilibrium states on Minkowski space, 
as we shall see. 

The two--point functions of the  
KMS--states are, for $x,y \in \cW$, also given by relation
(\ref{f7}), where one has to replace everywhere the quantities
on Minkowski space 
$({\cM}_0,g_0)$ by the corresponding ones on de Sitter space
$({\cM},g)$ and $t \mapsto x(t)$ is now given by relation
(\ref{f10}). The analogue of relation (\ref{f8}) is
\begin{equation*} 
\begin{split}
& \wb{\big(\phi_{(\cM,g)} (x)  \,  \phi_{(\cM,g)} (y)
- H_{(\cM,g)}(x,y) \big) } =  \\
& \text{\small $\frac{i}{2 \pi}$} \! \!   
\int \! \!  d\omega \, \!  
\Big( \text{\small $\frac{1}{1 - e^{-\beta \omega}}$} 
- \text{\small $\frac{1}{1 - e^{- \beta_r \omega}}$}  \Big) \! \! 
\int \! \!  dt  \Delta_{(\cM,g)}(x(t),y) \, e^{it\omega} 
- \text{\small $\frac{1}{768 \pi^2}$} \, R \, .
\end{split}
\end{equation*}
Again, one may proceed here to the limit 
$y \rightarrow x$. The undetermined constant $c$ in relation 
(\ref{f5}) can then be fixed with the help of the above
relation by imposing the condition that $\phi^2$, which macroscopically 
is proportional to the square of the temperature, 
has expectation value $0$ on the geodesic of the observer
in the ground state, \ie the KMS--state for $\beta \rightarrow \infty$. 
Thus, with $t \mapsto w(t)$ as above, 
\begin{equation*}
c - \text{\small $\frac{1}{768 \pi^2}$} 
= \lim_{y \rightarrow w}  \text{\small $\frac{i r^2}{12 \pi}$}  
\int_0^\infty \!   d\omega \, 
\text{\small $\frac{1}{e^{\beta_r \omega} - 1}$}  \!
\int \! dt \,  \Delta_{(\cM,g)}(w(t),y) \, e^{it\omega} ,
\end{equation*}
which, after a straightforward computation, yields for
any $r > 0$ the value 
\begin{equation*} 
c = \text{\small $\frac{1}{192 \, \pi^2}$} \, . 
\end{equation*}

We mention as an aside that the definition of 
the Wick square (\ref{f5}) has 
thereby been fixed on all spacetimes.
This method of specifying ``renormalization constants'' on 
the basis of a thermal interpretation of the composite 
fields seems to be of general interest.

In order to see whether a state $\w{\cdot}$ on 
$(\cM,g)$ has a thermal
interpretation with regard to $\phi_{(\cM,g)}^2$, one has  
to rely on relations (\ref{f3}) and (\ref{f9}), \ie one has to check whether
there exists some probability measure $\rho_x$ such that
\begin{equation*} 
\w{\phi_{(\cM,g)}^2(x)} = 
\int \! d\rho_x(\beta) \,  \text{\small $\frac{1}{12  \, \beta^2}$} \, .
\end{equation*}
Since we are dealing here with a single observable, this 
condition is satisfied if and only if the expectation value is 
positive \cite{remark}.
There then exists an abundance of measures 
solving this equation. What matters here is that we can interpret
under these circumstances the expectation value 
$\w{12 \, \phi_{(\cM,g)}^2(x)}$ as 
the (mean) square of the temperature at $x$,
which we denote by  $ T^2(x) = \text{\small $\frac{1}{\beta^2(x)}$} $.  

We use now the observable $\phi_{(\cM,g)}^2$ for the analysis
of the thermal properties of the KMS--states at arbitrary points
$x \in \cW$. Its expectation values are readily obtained from 
the above relations \cite{Sch}
\begin{equation*} 
\begin{split}
& \wb{\phi_{(\cM,g)}^2(x)} 
= \text{\small $\frac{r^2}{4 \pi^2 \, ({x_1}^2 - {x_0}^2) }$}  
\int \!   d\omega \, 
\Big( \text{\small $\frac{1}{1 - e^{- \beta \omega}}$} - 
      \text{\small $\frac{1}{1 - e^{- \beta_r \omega}}$} \Big)
\, \omega + (c - \text{\small $\frac{1}{768 \pi^2}$}) \, R  \\
& = \text{\small $\frac{r^2}{({x_1}^2 - {x_0}^2)}$}   \,
\Big(  \text{\small $\frac{1}{12 \, \beta^2}$} -
 \text{\small $\frac{1}{12 \, \beta_r^2}$} \Big) +
 \text{\small $\frac{1}{12 \, \beta_r^2}$} \, .
\end{split}
\end{equation*}
Hence the (mean) square of the 
local temperature in the KMS--states is given by
\begin{equation} \label{f11}
 T^2(x) =  
 \text{\small $\frac{r^2}{({x_1}^2 - {x_0}^2)}$}   \,
(  T^2 - T_r^2) + T_r^2 \, , \quad x \in \cW \, ,
\end{equation}
provided the right hand side is positive. It is constant
along the orbits (\ref{f10}). 
Restricting attention first to the points $w$ on the geodesics 
for which $w_1^2 - w_0^2 = r^2$,  we
find that $T^2(w) = T^2$. So, on the geodesics 
the parameter $\beta$ appearing in the KMS--condition indeed has the
physical significance of inverse temperature.
For points $x \in \cW$ off the geodesics one has $0 < x_1^2 - x_0^2 < r^2$,
where the lower bound is attained if one reaches the horizon. 
Thus unless $T = T_r$, \ie the KMS--state
is the Gibbons--Hawking state, $T^2(x)$ depends 
non-trivially on $x$. 
If  $T > T_r$ the temperature increases if one
departs from the geodesic and reaches arbitrarily high
values close to the horizon. For  $T < T_r$ the
temperature decreases if one leaves the geodesic. As a matter of fact,
the right hand side of relation (\ref{f11}) becomes
negative if $x$ is sufficiently close to the 
horizon. So a thermal interpretation of these states breaks 
down at such points, they are locally out of equilibrium. 

These results can be understood as a consequence of the 
Unruh effect. An observer moving along the worldline
(\ref{f10}) will find that with respect to his proper dynamics 
and time scale the parameter fixing the underlying KMS--state 
has the smaller value 
\begin{equation*}
\widetilde{\beta}(x) = \sqrt{\frac{{x_1}^2 - {x_0}^2}{r^2}} \, \beta \, ,
\end{equation*}
corresponding to a hotter environment at temperature 
$\widetilde{T}(x) =  
\text{\small $\frac{1}{\widetilde{\beta}(x)}$}$. Taking into
ac\-count that he also 
experiences a constant acceleration along his world line given~by

\vspace*{-3mm} 
\begin{equation*}
a(x) =  \sqrt{ \text{\small $\frac{1}{{x_1}^2 - {x_0}^2}$} 
         -  \text{\small $\frac{1}{r^2}$}} \, , 
\end{equation*}
one can rewrite equation (\ref{f11}) in the form

\vspace*{-3mm} 
\begin{equation} \label{f12} 
{\widetilde{T}(x)} =
\sqrt{ \, T^2(x)
+ \frac{a(x)^2}{4 \pi^2}} \, .
\end{equation}
Thus the temperature felt by this observer 
results from the local temperature $\sqrt{T^2(x)}$ along his
world line and the Unruh effect due to his acceleration $a(x)$.
Relation (\ref{f12}) generalizes a formula given in \cite{NaPeTh,DeLe,Ja}
for the Gibbons--Hawking state; note that it holds for 
all KMS states, irrespective of the value of $T$. We  
therefore conjecture that it is meaningful in all states 
on de Sitter space where a local temperature can be defined
in the sense explained above.

The preceding results show that it is possible
to introduce in quantum field theory on curved spacetime 
a local notion of temperature in an operationally 
meaningful manner. Every observer who uses the according
to relation (\ref{f5}) calibrated ob\-servable $\phi_{(\cM,g)}^2$ will 
find in a local equilibrium state at given spacetime point $x$
a definite value $\sqrt{T^2(x)}$ of the temperature,
irrespective of his motion. He simply has to 
determine the mean of $\phi_{(\cM,g)}^2(x)$ in this state 
\cite{remark2}. The resulting value has to be interpreted as 
the true local temperature of the system at $x$. In contrast,
the quantity $\widetilde{T}(x)$ has the meaning of the temperature in the 
laboratory system of the moving observer, which deviates 
from   $\sqrt{T^2(x)}$ if he is subject to acceleration. In other words, 
$\sqrt{T^2(x)}$ is the temperature felt by geodesic observers 
passing~through~$x$. 

\noindent{\Large \bf Acknowledgments} \\[4mm]
We would like to thank J.~Bros for communicating to
us his unpublished results on the commutator function
on de Sitter space.


\begin{thebibliography}{99}
\bibitem{remark}
Note that due to the Wick ordering (\ref{f5}), the expectation values 
of $\phi_{(\cM,g)}^2(x)$ 
are negative in many states.
\bibitem{remark2}
This requires, of course, that he is in a position to
repeat this experiment -- which is taken for granted in any
theory with a statistical interpretation.
\end{thebibliography}
\end{document}